\documentclass[aps,prb,twocolumn,floatfix]{revtex4}
\usepackage{amsmath}
\usepackage{graphicx}

\begin{document} 

\title{ A Rational Indicator of Scientific Creativity } 

\author{ Jos\'e M. Soler\footnote{E-mail: jose.soler@uam.es.} }
\affiliation{ Departamento de F\'{\i}sica de la Materia Condensada, 
              C-III, Universidad Aut\'{o}noma de Madrid,
              E-28049 Madrid, Spain }

\date{\today}

\begin{abstract}
  A model is proposed for the creation and transmission of scientific 
knowledge, based on the network of citations among research articles.
   The model allows to assign to each article a nonnegative value for
its creativity, i.\ e.\ its creation of new knowledge.
   If the entire publication network is truncated to the first
neighbors of an article (the $n$ references that it makes and the $m$
citations that it receives), its creativity value becomes a simple 
function of $n$ and $m$.
   After splitting the creativity of each article among its authors, 
the cumulative creativity of an author is then proposed as an 
indicator of her or his merit of research.
   In contrast with other merit indicators, this creativity index 
yields similar values for the top scientists in two very different 
areas (life sciences and physics), thus offering good promise for
interdisciplinary analyses.
\end{abstract}

\maketitle 

\section*{Introduction}

   Evaluating the scientific merit and potential, of tenure and
professorship candidates, is perhaps the most critical single activity 
in the academic profession.
   In countries and institutions with a long scientific tradition,
selection committees are generally well trained and trusted to balance 
wisely the vast variety of factors that may influence the decision, 
in the sense of optimizing the long-term scientific output.
   In less established environments, decisions are frequently perceived
as arbitrary, and the use of objective indicators and procedures may be
necessary to obtain a wide consensus.\cite{Moed2005}

   The most traditional indicator of research output, the number of 
published papers, has been progressively substituted by the number
of citations received by those papers, when this impact indicator 
has become widely available and easy to obtain.\cite{Garfield1964,ISI}
   Different combinations of both magnitudes have been proposed,
\cite{Rinia1998} like those in the SPIRES database.\cite{SPIRES}
   The field has been recently revitalized by the proposal by 
Hirsch~\cite{Hirsch2005} of yet another combination, the so-called
$h$ index, which has gained a rapid popularity, partly because the
Thomson-ISI Web of Knowledge database~\cite{ISI} provides a handy tool
to sort articles by their number of citations (while it offers no tools 
to obtain other indicators, like the total citation count).
   Apart from that comparative handiness, there is little objective 
evidence for the relative advantages of different indexes, which are 
generally motivated in terms of ``impact'' or ``influence''.
   However, it must not be forgotten that the task of a scientist is 
to create useful knowledge (in its broadest sense), not merely to 
produce an impact.
   It is therefore desirable to derive some rational measure of
the magnitude and quality of research output, rooted in a plausible 
model of the creation and transmission of scientific knowledge.
\cite{Rinia2002}

\section*{Creativity Model}

   Basic scientific knowledge, as opposed to technological or industrial
knowledge, is created by the minds of scientists and expressed almost
exclusively as research articles.
   The knowledge is transmitted to other scientists, who read previous
articles and acknowledge this transmission in the form of references
(in what follows, I will call {\em references} of an article
those made to previous papers, and {\em citations} those 
received from posterior papers).
   Thus, the output knowledge of an article comes partly from
previous work, which is simply transmitted, and partly from the
creation of new knowledge by the authors.
   However, there are many possible reasons why references are made.
\cite{Garfield1964,Merton1968,Gilbert1977,Cozzens1989}
   Furthermore, some of the references of an article may be more 
important than others.
   Thus, it is rather uncertain to what extent a given reference 
reflects the use of previous knowledge.
   Therefore, in the present model I will simply assume that each 
reference reflects the transmission of a different nonnegative value 
$x_{ij}$ of knowledge, with probability $P(x_{ij})$, from the cited 
article $i$ to the citing article $j$.
   The maximum entropy principle~\cite{Tribus1969} dictates that,
in the absence of any {\it a priori} information, other than the 
average value $\langle x \rangle = 1/\alpha$, the probability is given 
by $P(x)=\alpha e^{-\alpha x}$.

   Consider the network formed by all published papers connected by 
their citations.
   The growth, connectivity, and statistical properties of this and 
similar networks have been the subject of much recent work.
\cite{Redner2005,Albert-Barabasi2002}
   To model the flow of knowledge on this supporting network,
\cite{Rinia2002} we may assign random flow numbers $x_{ij}$ to all 
citations, with probability $P(x_{ij})$.
   Flow conservation implies that the articles' knowledge-creation 
values $c_i$ (that I will simply call creativities) obey
%------------------------------------------------------------------
\begin{equation}
   c_i = \sum_j x_{ij} - \sum_k x_{ki}
\label{ci}
\end{equation}
%------------------------------------------------------------------
   I will discard negative knowledge as 
meaningless.\cite{negative_c}
   Thus, I will require that $c_i \ge 0 ~~\forall i$, and reject the 
sets $\{x_{ij}\}$ that violate this condition.\cite{alpha,gamma,boundary}
   The final values $c_i$ will then be averages over all valid sets
$\{x_{ij}\}$, with a relative weight 
$P(\{x_{ij}\}) \propto \exp(-\alpha \sum_{ij} x_{ij})$.

   Some attention must be paid to the definition of knowledge that 
is being used.
   It might seem that all the knowledge created by an article must 
be present already when it is published.
   However, this would make it difficult to judge the relative 
importance of the knowledge created by different papers.
   Therefore, I rather consider the amount of ``used knowledge''
(and therefore useful).
   The situation is very similar in software development:
the economic value of a computer library does not materialize when it 
is written, but when licenses of it are sold, presumably to create 
new software (for free software we might substitute licenses sold 
by copies downloaded).
   Similarly, I am counting every ``copy'' of the knowledge, used in 
every new paper that cites it
(alternatively, one might consider the knowledge created by a paper
as the sum of that added to all the brains that have read it).

   Some of the general qualitative features of the model, as an 
indicator of research merit, may be expected {\it a priori}:
   articles with less citations than references will have a positive
but small creativity value;
   articles with a large output (very cited) and a small input (not 
many references) will have the largest creativities;
   in contrast, the merit of review articles will be much more moderate
than that shown by their raw impact factor (citation count);
   the differences between the creativities of authors in very large and 
active fields (with large publication and citation rates), and those in
smaller and less active fields, will be largely attenuated, as compared 
to other merit indicators, since the basic measure is the difference 
between citations and references, which should be roughly zero in all 
fields;
   self-citations will be largely discounted, since they will count
both as a negative contribution (to the citing paper) and a positive 
one (to the cited paper);
   citations received from a successful article (i.\ e.\ a very cited
one itself) will be more valuable than those made by a poorly cited one.
\cite{Pinski-Narin1976,Chen2006}
   In particular, citations by uncited papers will add no value at all,
since no knowledge can flow through them;
   more generally, articles that generate a divergent citation tree
(e.\ g.\ the DNA paper of Watson and Crick) will have a large creativity,
while those leading ultimately to a dead end (e. g. the cold fusion 
paper of Fleischmann and Pons) will have a small one, even if they
had the same number of direct citations.

\section*{Simplified Model}

\begin{figure}[htpb]
\includegraphics[width=0.9\columnwidth,clip]{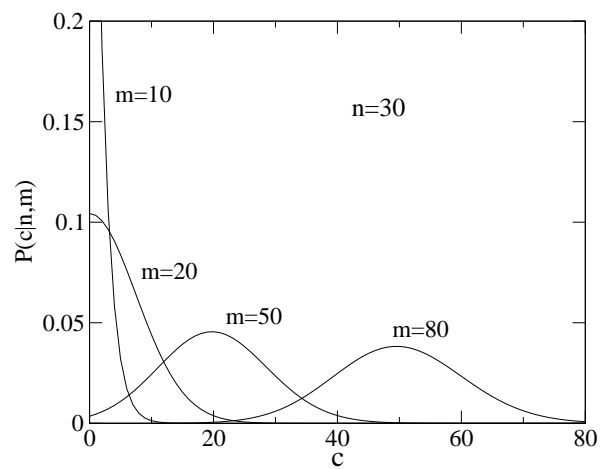}
\caption{
   Probability that an article, that has made $n=30$ references and has 
received $m$ citations, has created a value $c$ of scientific
knowledge. It was obtained from Eq.~(\ref{Pexact}).
}
\label{Pfig}
\end{figure}

   The quantitative analysis of the model presented above is an 
interesting challenge that will be addressed in the future.
   In this work, I am rather interested in simplifying the model to 
allow the easy generation of a practical indicator of merit
of research.
   The simplified model will keep many of the general features 
discussed  above, though not all (in particular, it will loose the 
last two properties mentioned above).
   Thus, I propose to truncate the citation network beyond the
first neighbors of any given paper, i.\ e.\ to consider only its
$n$ references and $m$ citations, and to impose the conservation
of flow, Eq.~(\ref{ci}), only in the central node $i$.
%   To simplify the model to its bare minimum, I will also ignore
%hidden flow ($\gamma=0$) and the change of average flow per
%citation ($\alpha=$~const).
   The average value $\langle x \rangle$ can be used as a convenient
unit of knowledge, so that $\alpha=1$ and $P(x)=e^{-x}$.
   The probability that an article, with $n$ references
and $m$ citations, has a creativity $c$ is then, for $n,m>0$:
%------------------------------------------------------------------
\begin{equation}
P(c|n,m) = N^{-1} \int...\int_0^\infty dx_1...dx_n dy_1...dy_m
  ~\delta(c + x - y) ~e^{-x-y}
\label{Pnmc}
\end{equation}
%------------------------------------------------------------------
with $x=\sum_{i=1}^n x_i$ and $y=\sum_{j=1}^m y_j$, where $x_i$ are 
the input flows (references) and $y_j$ are the outputs (citations).
   $\delta(x)$ is Dirac's delta function, and $N$ is a 
normalization factor given by
%------------------------------------------------------------------
\begin{equation}
N = \int...\int_0^\infty dx_1...dx_n dy_1...dy_m 
     ~\theta(y-x) ~e^{-x-y}
\end{equation}
%------------------------------------------------------------------
where $\theta(x)$ is the step function.
   Using a convenient change of variables, the integrals can be
evaluated as
%------------------------------------------------------------------
\begin{equation} 
N = \int\int_0^\infty  
    ~\frac{dx ~dy ~x^{n-1} y^{m-1}}{(n-1)!(m-1)!} 
    ~\theta(y-x) ~e^{-x-y}
\end{equation}
%------------------------------------------------------------------
\begin{equation}
P(c|n,m) 
  = N^{-1} \int\int_0^\infty 
    ~\frac{dx ~dy ~x^{n-1} y^{m-1}}{(n-1)!(m-1)!}
    ~\delta(c+x-y) ~e^{-x-y}
\end{equation}
%------------------------------------------------------------------
   The result is
%------------------------------------------------------------------
\begin{equation}
P(c|n,m) 
= \frac{n e^{-c}}{n+m-1} ~\frac{_1F_1(1-m,2-n-m;2c)}{_2F_1(1,1-m;1+n;-1)}
\label{Pexact}
\end{equation}
%------------------------------------------------------------------
where $_1F_1$ and $_2F_1$ are hypergeometric functions, which can
be expanded as a finite series.\cite{Gradshteyn-Rydhik1980}
   Figure~\ref{Pfig} shows some typical probability distributions.
%=====================================================================
%=====================================================================

\begin{figure}[htpb]
\includegraphics[width=0.9\columnwidth,clip]{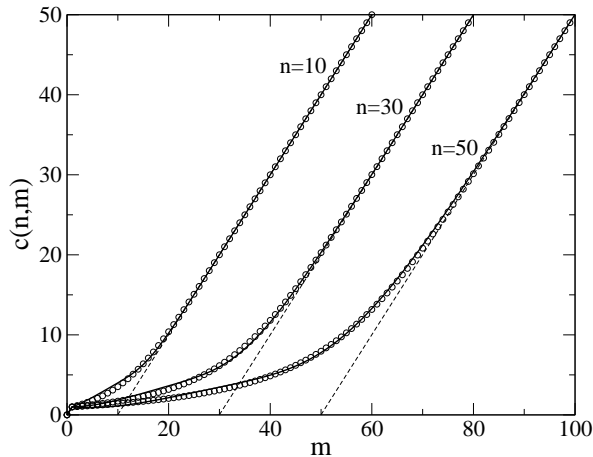}
\caption{
   Circles: mean creation of knowledge (creativity) of an article with 
$n$ references and $m$ citations, calculated from Eq.~(\ref{cnm})
(in units of the mean transmission of knowledge reflected by one 
reference).
   Solid lines: fits given by Eq.~(\ref{cfit}).
   Dashed lines: $m-n$.
}
\label{cfig}
\end{figure}

   The average value of $c$,
%------------------------------------------------------------------
\begin{equation}
c(n,m) = \int_0^\infty dc ~c ~P(c|n,m),
\label{cnm}
\end{equation}
%------------------------------------------------------------------
is, for $n,m>0$:
%------------------------------------------------------------------
\begin{equation}
c(n,m) =
 \frac{ \sum_{k=0}^{m-1} \frac{(n+m-2-k)!}{(m-1-k)!} (k+1) 2^k }
      { \sum_{k=0}^{m-1} \frac{(n-1)!(n+m-1)!}{(n+k)!(m-1-k)!} }.
\end{equation}
%------------------------------------------------------------------
   It is represented in figure~\ref{cfig} for some typical values
of $n$ and $m$.
%=====================================================================
%=====================================================================
   As expected, $c(n,m)$ increases with $m$ and it decreases with $n$.
   It obeys $c(0,m)=m$, $c(n,0)=0$, $c(n,1)=1$, and 
$c(n,m) \ge \max(1,m-n) ~~\forall m>0$.
   For the present purposes, a reasonably accurate fit is, for $m>0$:
%------------------------------------------------------------------
\begin{equation}
c(n,m) \simeq m - n + \frac{n}{ A ~e^{a z} + B ~e^{b z} }
\label{cfit}
\end{equation} 
%------------------------------------------------------------------
where $z=(m-1)/(n+5)$, $A=0.986, B=0.014, a=1.08$, and $b=6.3$.
   The accumulated creativity of an author with $N_p$ published papers
is then defined as 
\begin{equation}
C_a = \sum_{i=1}^{N_p} \frac{c(n_i,m_i)}{a_i} 
\label{Cauthor}
\end{equation} 
where $a_i$ is the number of authors of paper $i$.
   Notice that, being positive and cumulative, $C_a$ can only increase
with time and with the number of published papers.

   In order to find in practice the creativity of an author (among
many other merit indicators), one can follow these steps:
%\begin{enumerate}
%\item
   1)
   Download the programs {\it filter} and {\it merit} from this
   author's web page,~\cite{soler_web} and compile them if necessary.
%\item
   2)
   Perform a ``General search'' in the Thomson ISI Web of Science 
   database~\cite{ISI} for the author's name, using the appropriate
   filters.
%\item
   3)
   Select the required records. Usually the easiest way is to check
   ``Records from 1 to last\_one'' and click on ``ADD TO MARKED LIST''
   (if you find too many articles, you may have to mark and 
   save them by parts, say (1-500)$\rightarrow$file1, 
   (501-last\_one)$\rightarrow$file2);
%\item
   4)
   Click on ``MARKED LIST''.
%\item
   5)
   Check the boxes ``Author(s)'', ``Title'', ``Source'', ``keywords'',
   ``addresses'',  ``cited reference count'', ``times cited'',
   ``source abbrev.'', ``page count'', and ``subject category''.
   Do not check ``Abstract'' nor ``cited references'', since this
   would slow down considerably the next step.
%\item
   6)
   Click on ``SAVE TO FILE'' and save it in your computer.
%\item
   7)
   Click on ``BACK'', then on ``DELETE THIS LIST'' and ``RETURN'',
   and go to step 2 to make another search, if desired.
%\item
   8)
   If you suspect that there are two or more authors with the same
   name, use the {\it filter} program to help in selecting the papers
   of the desired author.
%\item
   9)
   Run the {\it merit} program to find the merit indicators.
   Mind for hidden file extensions, possibly added by your navigator, 
   when giving file names in this and previous step.
%\end{enumerate}

\section*{Results and Discussion}

   Table \ref{Phys&Biol} shows several indexes of merit of top
scientists in life sciences and physics, taken from Hirsch's 
selection.\cite{Hirsch2005}
%=====================================================================
\begin{table}
\begin{center}
\begin{tabular}{cccccc}
Name            &  $N_p$ & $N_c (10^3)$ & $h$ & $P_a (10^3)$& $C_a (10^3)$ \\
\hline
B. Vogelstein   &     447 &    144.4 &     154 &     34.1 &     32.0 \\
S. H. Snyder    &    1144 &    138.3 &     194 &     48.2 &     38.9 \\
S. Moncada      &     693 &    106.2 &     145 &     32.5 &     27.8 \\
P. Chambon      &     987 &     98.1 &     153 &     23.0 &     17.7 \\
R. C. Gallo     &    1247 &     95.9 &     154 &     17.9 &     13.8 \\
D. Baltimore    &     657 &     95.3 &     162 &     33.0 &     28.2 \\
R. M. Evans     &     428 &     78.8 &     130 &     21.2 &     18.3 \\
T. Kishimoto    &    1621 &     77.5 &     134 &     14.6 &     10.2 \\
C. A. Dinarello &     992 &     74.3 &     138 &     26.3 &     19.2 \\
A. Ullrich      &     615 &     73.0 &     122 &     13.6 &     10.9 \\
{\bf Average}   &{\bf 883}&{\bf 98.2}&{\bf 149}&{\bf 26.4}&{\bf 21.7} \\
Standard dev.   &     364 &     24.1 &      19 &     10.1 &      9.1 \\
\hline
P. W. Anderson  &     342  &    56.7 &     96 &      39.1 &     36.9 \\
A. J. Heeger    &     999  &    53.5 &    109 &      14.2 &     10.3 \\
E. Witten       &     254  &    53.1 &    111 &      39.9 &     35.9 \\
S. Weinberg     &     444  &    38.8 &     88 &      32.7 &     29.3 \\
M. L. Cohen     &     625  &    37.4 &     94 &      14.3 &     10.6 \\
M. Cardona      &    1096  &    37.0 &     88 &      12.8 &      7.8 \\
A. C. Gossard   &     918  &    34.3 &     92 &       7.4 &      5.8 \\
P. G. deGennes  &     358  &    32.6 &     80 &      26.7 &     23.9 \\
M. E. Fisher    &     446  &    29.8 &     88 &      19.0 &     14.3 \\
G. Parisi       &     469  &    24.9 &     75 &      12.2 &      9.9 \\
{\bf Average}   &{\bf 595}&{\bf 39.8}&{\bf 92}&{\bf  21.8}&{\bf 18.5}\\
Standard dev.   &     286  &    10.4 &     11 &      11.3 &     11.3 \\
\hline
%R. Jackiw       &     197  &    22.6 &     70 &     10.0 \\
%Z. Fisk         &     830  &    21.9 &     76 &      2.6 \\
%J. N. Bahcall   &     421  &    20.9 &     76 &      5.4 \\
%F. Wilczek      &     307  &    20.5 &     69 &      8.9 \\
%%D. J. Gross     &     132  &    20.1 &     62 &      8.1 \\
%S. W. Hawking   &     138  &    19.0 &     61 &     12.7 \\
%S. G. Louie     &     325  &    18.6 &     75 &      4.2 \\
%D. J. Scalapino &     371  &    18.1 &     76 &      4.9 \\
%M. B. Maple     &     703  &    16.7 &     67 &      2.0 \\
%C. Vafa         &     135  &    13.6 &     67 &      4.5 \\
%%M. S. Dresselhaus &   400  &    11.2 &     52 &      1.4 \\
\end{tabular}
\end{center}
\caption{
   Several merit indicators of the ten most cited scientists in life
sciences and physics.\cite{Hirsch2005}
   $N_p$: number of papers published. 
   $N_c$: number of citations received by those papers. 
   $h$: number of papers with $h$ or more citations 
(Hirsch index).\cite{Hirsch2005}
   $P_a$: author's knowledge-productivity index, 
$P_a=\sum_{i=1}^{N_p} m_i/a_i$, 
where $a_i$ and $m_i$ are the number of authors and of citations 
received by paper $i$.
   $C_a$: author's creativity index, Eq.~(\ref{Cauthor}).
   The data were obtained in April 2006.
}
\label{Phys&Biol}
\end{table}
%=====================================================================
   It may be seen that the $h$ index of all biologists 
is larger than that of all physicists, and their average number of 
publications and citations is 1.5--2.5 times larger.
   In contrast, the two creativity distributions are remarkably 
similar, with averages that differ only $\sim 15\%$, well below 
the standard deviation of both distributions.
   This offers the promise of direct interdisciplinary comparisons,
without any field normalization, a highly desirable characteristic 
of any index of merit.
%    It also renders credibility to the model, since it is reasonable
% to expect that the top scientists of two large areas like biology
% and physics should have similar creativities.

   Although it is a natural consequence of the idea of knowledge flow, 
the fact that the references of an article will result in lowering 
the merit assigned to it, is admittedly striking.
   It is thus appropriate to recognize that this is partly due to 
a deliberate intent of measuring creativity rather than productivity
(or, in economic terms, added value rather than sales).
   To illustrate the point, imagine that two scientists, Alice and
Bob, address independently an important and difficult problem in 
their field.
   Bob takes an interdisciplinary approach and discovers that a
method developed in a different field just fits their need.
   Simultaneously, Alice faces the problem directly and re-invents
the same method by herself (thus making less references in her
publication).\cite{good_practices}
   All other factors being equal, both papers will receive roughly 
the same number of citations, since they transmit the same knowledge 
to their field.
   But it may be argued that Alice's work was more creative in some 
sense, and that her skills might possibly (but not necessarily)
be more valuable in a given selection process.
   Eventually, the usefulness of different merit indicators will
depend on how well they correlate with real human-made selections
\cite{Cole1971,Rinia1998}.
   Thus, Table~\ref{Phys&Biol} shows also a ``productivity index'' 
$P_a$ (not a probability), given by the author's share of the 
citations received by her/his papers.
   Notice that, in the model proposed, $N_c$ is the total output 
flow of knowledge from the author's papers, while $P_a$ is her/his 
share of it.
   It may be seen that $P_a$ also allows reliable interdisciplinary 
comparisons.
   It may be concluded that the main difference between the two
communities is the larger average number of authors per article in 
the life sciences, which is taken into account in both $P_a$ and $C_a$,
but not in the other indexes.
%    A small random survey among US university professors in physics
% and biology showed that typical (median) values for creativity 
% (productivity) are $\sim 300 (600)$ for full professors and 
% $\sim 50 (100)$ for associate and assistant professors, 
% but with very large variances.

   Knowledge-productivity and creativity indicators can be used also 
for groups, institutions, or journals.
   Thus, Table~\ref{physics_journals} shows them for some leading 
journals.
%=====================================================================
\begin{table}
\begin{center}
\begin{tabular}{cccccc}
Journal & $N_p$ & $N_r/N_p$ & $N_c/N_p$ & $C/N_p$ & IF \\
\hline
Nature            & 3676 &  10 &  67 &  59 & 28.8 \\
Science           & 2449 &  14 &  74 &  63 & 24.4 \\
\hline
Rev. Mod. Phys.   &   20 & 284 & 327 & 160 & 13.4 \\
Adv. Phys.        &    8 & 391 & 149 &  18 & 12.7 \\
Surf. Sci. Rep.   &    5 & 159 &  61 &   3 & 10.3 \\
Rep. Prog. Phys.  &   29 & 198 &  90 &  32 &  6.2 \\
Phys. Rep.        &   81 & 166 &  90 &  22 &  5.6 \\
\hline
Phys. Rev. Lett.  & 1904 &  18 &  59 &  44 &  6.0 \\
Phys. Rev. D      & 1049 &  27 &  23 &  11 &  3.9 \\
Nucl. Phys. B     &  620 &  37 &  42 &  24 &  3.3 \\
Appl. Phys. Lett. & 1819 &  13 &  34 &  26 &  3.3 \\
J. Chem. Phys.    & 2040 &  37 &  37 &  16 &  3.1 \\
Phys. Rev. B      & 3488 &  27 &  35 &  18 &  2.8 \\
\hline
\end{tabular}
\end{center}
\caption{
   Several indicators of some of the main multidisciplinary,
review and non-review Physics journals.
   $N_p$: number of ``papers'' (documents) published in year 1990, 
in all the sections included in the Science Citation Index 
database.
   $N_r$: number of references made by those papers. 
   $N_c$: number of citations received by those papers until May 2006. 
   $C$: Sum of the creativities, Eq.~(\ref{cnm}), of those papers,
$C = \sum_{i=1}^{N_p} c(n_i,m_i)$.
   IF: Impact factor in 1998 (center of the period 1990-2006), 
as defined by the Journal of Citation Reports.\cite{ISI}
   For the non-review physics journals (last group), the indicators 
(other than $N_p$ and IF) have been obtained from a random sample of 
their $N_p$ papers, rather than from the whole set.
}
\label{physics_journals}
\end{table}
%=====================================================================
   As expected, most review journals have considerably smaller 
creativities than productivities (dramatically smaller in some cases).
   Still, {\it Reviews of Modern Physics} has the largest creativity 
index of all the journals studied, showing that collecting, processing, 
and presenting knowledge in a coherent way can by itself create much 
{\em new} useful knowledge.

   Finally, in a world of strong competition for positions and founds,
a negative merit assignment to references might result in a 
tendency to reduce them below what would be scientifically desirable 
and professionally fair.
   A possible solution is to use, in Eq.~(\ref{cnm}), a fixed value of 
$n$ (equal to the journal reference intensity, i.\ e.\ the average number 
of references per article in that journal), to calculate the creativities 
for competitive-evaluation purposes.
   This would spoil a few desirable properties of the model (like 
the discount of self-citations), but most of its effects would probably 
be rather mild, since the number of references per paper has a much 
smaller variance than the number of citations.
   Thus, the root mean squared difference between the creativities of 
Table~\ref{Phys&Biol}, calculated using the average references
of the journals, rather than the actual references of each article,
is only $\sim 4\%$.

\section*{Conclusion}

   In conclusion, I have proposed an index of research merit based
on creativity, defined as the creation of new scientific knowledge, 
in a plausible model of knowledge generation and transmission.
   It is calculated easily from the citations and references of the
author's articles, and it is well suited for interdisciplinary 
comparisons.
   An advantage of such an index is that its meaning may be more 
easily perceived, by policy makers and the general public, as a 
measure of a scientist's social and economic service to the community.

\begin{acknowledgements}
   I would like to acknowledge very useful discussions with
J. V. Alvarez, J. R. Castillo, R. Garc\'{\i}a, J. G{\'o}mez-Herrero, 
L. Seijo, and F. Yndurain.
   This work has been founded by Spain's Ministery of Science through
grant BFM2003-03372.
\end{acknowledgements}

\bibliographystyle{apsrev}
\bibliography{creativity}

\begin{thebibliography}{24}
\bibitem[{neg()}]{negative_c}
\bibinfo{note}{An ironic observer might object to this assumption, arguing that
  many articles contribute only to confusion, and that some citations are in
  fact critical. I find this questionable, since most readers will filter
  efficiently this ``negative'' knowledge, simply ignoring it. Also, even wrong
  ideas can stimulate new valid ones. In any case, critical references cannot
  be easily distinguished from positive ones, but their average effect might be
  taken into account by renormalizing the mean flow value $\langle x \rangle$.}
\bibitem[{alp()}]{alpha}
\bibinfo{note}{Since new, nonnegative knowledge is created in every article and
  transmitted to the future, the total flow of knowledge must increase with
  time. Such an increase may be absorbed in three ways: by an increase in the
  number of articles published per year; by an increase in the number of
  references per article; and by an increase in the average flow per citation
  $\langle x \rangle$. The increase of the rate of publications is indeed a
  large effect, while that of citations per paper is much weaker, if positive
  at all. In any case, it is not clear whether those two effects combined can
  fully account for the transmission of the new knowledge predicted by the
  model. Thus, it may be necessary to adjust self-consistently a function
  $\alpha(t)$, of time $t$. In this work I have taken $\alpha$ = const = 1.}
\bibitem[{bou()}]{boundary}
\bibinfo{note}{The boundary problem posed by recent papers, that have had no
  time to transmit their knowledge, may be addressed by not imposing flow
  conservation on them, or by assigning to them an average number of additional
  expected citations, that will be a decreasing function of their age. In any
  case, it is clear that any figure of merit based on citations will not be as
  reliable for very recent papers as for old ones.}
\bibitem[{goo()}]{good_practices}
\bibinfo{note}{This is somewhat hypothetical since good citation practice
  (generally enforced by the referees) requires that previous relevant work is
  cited, independently of whether it was actually used.}
\end{thebibliography}

\end{document}